\documentclass[10pt,conference]{IEEEtran}

\usepackage{amsmath,amsfonts,amssymb}
\usepackage{mathtools}
\usepackage{amsthm}
\usepackage{alltt, xspace, times, epsfig, url}
\usepackage{gensymb,xfrac,array,booktabs,bm,xcolor}
\usepackage{adjustbox}
\usepackage{booktabs}
\usepackage{soul}  
\usepackage{algorithm}
\usepackage{tabularx}
\usepackage{colortbl}

\usepackage{algpseudocode}

\def\BibTeX{{\rm B\kern-.05em{\sc i\kern-.025em b}\kern-.08em
    T\kern-.1667em\lower.7ex\hbox{E}\kern-.125emX}}

\usepackage{pbalance}

\makeatletter
\newcommand{\linebreakand}{%
  \end{@IEEEauthorhalign}
  \hfill\mbox{}\par
  \mbox{}\hfill\begin{@IEEEauthorhalign}
}
\makeatother

\DeclareMathOperator*{\argmax}{argmax}
\DeclareMathOperator*{\argmin}{argmin}
\theoremstyle{definition}

\begin{document}
\title{Defending Against Beta Poisoning Attacks in Machine Learning Models}

\author{%
\IEEEauthorblockN{Nilufer Gulciftci}
\IEEEauthorblockA{\textit{Department of Computer Engineering} \\
\textit{Acıbadem University}\\
Istanbul, Turkiye \\
nilufer.gulciftci@live.acibadem.edu.tr}
\and
\IEEEauthorblockN{M. Emre Gursoy}
\IEEEauthorblockA{\textit{Department of Computer Engineering} \\
\textit{Koç University}\\
Istanbul, Turkiye \\
emregursoy@ku.edu.tr}
}

\maketitle

\begin{abstract}
Poisoning attacks, in which an attacker adversarially manipulates the training dataset of a machine learning (ML) model, pose a significant threat to ML security. Beta Poisoning is a recently proposed poisoning attack that disrupts model accuracy by making the training dataset linearly nonseparable. In this paper, we propose four defense strategies against Beta Poisoning attacks: kNN Proximity-Based Defense (KPB), Neighborhood Class Comparison (NCC), Clustering-Based Defense (CBD), and Mean Distance Threshold (MDT). The defenses are based on our observations regarding the characteristics of poisoning samples generated by Beta Poisoning, e.g., poisoning samples have close proximity to one another, and they are centered near the mean of the target class. Experimental evaluations using MNIST and CIFAR-10 datasets demonstrate that KPB and MDT can achieve perfect accuracy and F1 scores, while CBD and NCC also provide strong defensive capabilities. Furthermore, by analyzing performance across varying parameters, we offer practical insights regarding defenses' behaviors under varying conditions. 
\end{abstract}


\section{Introduction} \label{sec:introduction}

Machine learning (ML) models have become integral components in various domains, including finance, healthcare, cybersecurity, and autonomous systems. However, the robustness and trustworthiness of ML models are frequently challenged by adversarial attacks \cite{kumar2020adversarial}. Poisoning attacks constitute an important category of adversarial attacks, in which an attacker purposefully manipulates the training dataset to compromise the integrity of an ML model, e.g., degrade model accuracy or mislead its predictions \cite{kumar2020adversarial,cina2024machine,tian2022comprehensive}. 

A wide range of poisoning attacks have been proposed in the literature \cite{biggio2012poisoning, liu2017robust, jagielski2018manipulating, munoz2017towards, huang2020metapoison, cina2021hammer, aghakhani2021bullseye}. Among them, this paper focuses on Beta Poisoning attacks \cite{cina2021hammer}. Unlike a traditional bilevel optimization-based formulation, Beta Poisoning employs a heuristic approach to craft poisoning samples, which makes the training dataset linearly nonseparable. This strategy significantly reduces the computational overhead of the attack while maintaining high effectiveness, especially against linear ML models \cite{cina2021hammer, kara2022beta}.

While there exist several general-purpose defenses against poisoning attacks in the literature, they do not specialize in defending against Beta Poisoning. In this paper, we propose four defense strategies tailored to specifically counter Beta Poisoning. Our defenses are motivated by our analysis and observations of distinguishing characteristics of poisoning samples generated by the Beta Poisoning attack. In particular, we observed that many poisoning samples have close proximity to one another, and they are centered around the mean of the target class, which is typically located far from the mean of a non-target class.

Leveraging these insights, we developed four defenses: kNN Proximity-Based Defense (KPB), Neighborhood Class Comparison (NCC), Clustering-Based Defense (CBD), and Mean Distance Threshold (MDT). KPB identifies the k-nearest neighbors of each sample and uses the close proximity observation, i.e., if the sample's average distance to its nearest neighbors is low, then the sample is likely to be a poisoning sample. NCC compares the majority class of the sample's nearest neighbors and relatively distant neighbors. CBD utilizes the observation that poisoning samples form a tight cluster near the mean of the target class, which is typically far from the mean of the non-target class. Finally, MDT uses a similar observation to CBD; however, it relies on a distance threshold rather than a clustering-based approach. 

To evaluate the effectiveness of our defenses, we conducted an experimental evaluation following the experimental setup of prior works \cite{cina2021hammer, kara2022beta}. We measured the accuracy, precision, recall, and F1-scores of our defenses using the CIFAR-10 and MNIST datasets. We observed that KPB and MDT reach perfect 1.0 scores in all metrics and datasets. CBD also reaches 1.0 scores on the MNIST dataset and close to 1.0 scores on CIFAR-10. NCC falls slightly behind the other defenses, mostly due to its lower precision. Overall, given the scores, we conclude that our defenses are quite effective in defending against Beta Poisoning. In addition, experiments conducted by varying the parameters of the defenses and by visualizing the defense outputs enable us to gain additional insights regarding how defense parameters should be chosen, and when and why the defenses become more effective.

\section{Related Work} \label{sec:relatedwork}

Data poisoning attacks, in which the attacker injects maliciously constructed training samples to damage the performance of the model, are a major threat to the security of machine learning models \cite{biggio2012poisoning, cina2023wild,jia2022certified, paudice2018detection}. Biggio et al. \cite{biggio2012poisoning} proposed one of the first works on bilevel optimization-based poisoning. Similar optimization formulations were applied to attack feature selection in \cite{frederickson2018attack, xiao2015feature} and linear regression \cite{liu2017robust, jagielski2018manipulating}. In fact, most poisoning attacks require solving a bilevel optimization problem to identify the optimal poisoning samples that maximize attack impact \cite{munoz2017towards, mei2015using}. However, it is costly to solve bilevel problems \cite{biggio2012poisoning, cina2023wild}. Therefore, Mei and Zhu \cite{mei2015using} used machine teaching and Krush-Kahn-Tucker (KKT) conditions for poisoning. Munoz-Gonzalez et al. \cite{poisoning_attack_1} used back-gradient optimization, aiming to attack a wider class of ML models and multi-class classification. MetaPoison \cite{huang2020metapoison} used a first-order method to approximate the bilevel problem via meta-learning. Geiping et al. \cite{geiping2020witches} aimed to make attacks less expensive and more visually imperceptible, and proposed a method based on gradient matching. The feature collision strategy was proposed in \cite{shafahi2018poison}, with the aim of creating poisoning samples that collide with target test samples in the feature space. To make attacks more transferable and applicable to multiple models, \cite{aghakhani2021bullseye} and \cite{zhu2019transferable} proposed to optimize poisoning samples on ensemble models.

The threat of data poisoning attacks has also fueled the development of defenses. One practical defense strategy is to detect poisoning samples using outlier detection \cite{paudice2018detection}. Steinhardt et al. \cite{steinhardt2017certified} proposed a defense technique that detects and cleans outliers while applying empirical risk minimization. In contrast, \cite{fan2022survey} concentrated on detecting contaminated samples via statistical biases or distance-based measures. The defense designed by Jin et al. \cite{jin2021incompatibility} aims to detect backdoor attacks by forming clusters based on the incompatibilities of the data. The defense presented by Peri et al. \cite{peri2020deep} improves k-nearest neighbors by incorporating deep learning models to detect and counteract clean-label poisoning attacks. Similarly, Jia et al. \cite{jia2022certified} analyze the inherent robustness of KNN classifiers against poisoning attacks by establishing theoretical guarantees. 

Most closely related to our work are Beta Poisoning attacks, proposed in \cite{cina2021hammer}. Kara et al.~\cite{kara2022beta} examined the effects of Beta Poisoning on linear and non-linear models, demonstrating its effectiveness on linear models. In this paper, we propose defenses specifically targeting Beta Poisoning, which differs from traditional poisoning methods in terms of the heuristic attack strategy it utilizes \cite{cina2021hammer}. Thus, unlike previous outlier detection, k-nearest neighbor, or clustering-based defenses, our defenses utilize the specific behavior and insights derived from Beta Poisoning attacks. This targeted nature of our defenses enables us to achieve high accuracy and F1-scores.

\section{Background and Preliminaries} \label{sec:background}

\subsection{Supervised Learning}

Beta Poisoning attacks target supervised machine learning models. In a supervised learning setting, let $\mathcal{X} \subseteq \mathbb{R}^d$ denote the feature space and $\mathcal{Y}$ denote the label space. The training dataset is denoted by $\mathcal{D}_{tr} = \{(x_1,y_1), (x_2,y_2), ... \}$, where $x_i \in \mathcal{X}$ and $y_i \in \mathcal{Y}$, and the validation dataset is denoted by $\mathcal{D}_{val}$. For $(x_i,y_i)$, we use the terms ``sample" and ``point" interchangeably. A supervised machine learning model $f_{\theta}: \mathcal{X} \rightarrow \mathcal{Y}$ with parameters $\theta$ is trained using $\mathcal{D}_{tr}$ with the goal of minimizing the loss $\mathcal{L}(\mathcal{D}_{tr}, \theta)$.

\subsection{Poisoning Attacks} \label{sec:poisoning}

Data poisoning presents an important threat in machine learning, since attackers may purposefully manipulate the training dataset to compromise model integrity \cite{kumar2020adversarial,cina2024machine,tian2022comprehensive}. In a data poisoning attack, the attacker injects one or more poisoning samples to $\mathcal{D}_{tr}$. When the model $f_{\theta}$ is trained on the poisoned $\mathcal{D}_{tr}$, the model also becomes poisoned. 

Let $x_p \in \mathcal{X}$ be a poisoning sample with label $y_p \in \mathcal{Y}$. The attacker wants to maximize the impact of $x_p$ towards achieving his/her goal. Typically, this can be formulated using a bilevel optimization problem \cite{biggio2012poisoning, cina2021hammer}:
\begin{alignat}{3}
    &\max_{x_p} \qquad &&\mathcal{L}(\mathcal{D}_{val}, \theta^*) \label{eq:outer_optim} \\ 
    & \;\;\text{s.t.} &&\theta^* \in \argmin_{\theta} \mathcal{L}(\mathcal{D}_{tr} \cup (x_p, y_p), \theta) \label{eq:inner_optim} \\
    & &&\mathbf{x}_{lb} \preceq x_p \preceq \mathbf{x}_{ub} \label{eq:bilevel_bounds}
\end{alignat}
Here, Equation \ref{eq:outer_optim} corresponds to the goal of the attacker. In the example above, the goal is to maximize the loss over $\mathcal{D}_{val}$; hence, this is an untargeted attack. Since the crafted sample $x_p$ is added to $\mathcal{D}_{tr}$, it affects the training process of the model, which is captured by Equation \ref{eq:inner_optim}. The training goal shifts from finding the optimal parameters $\theta^*$ which minimize $\mathcal{L}(\mathcal{D}_{tr}, \theta)$ to finding the optimal parameters $\theta^*$ which minimize $\mathcal{L}(\mathcal{D}_{tr} \cup (x_p, y_p), \theta)$. Lower and upper bounds ($\mathbf{x}_{lb}$, $\mathbf{x}_{ub}$) are specified in Equation \ref{eq:bilevel_bounds} to constrain the search space for $x_p$.

It is important to highlight that the model $\theta^*$ is trained on the poisoned training dataset and subsequently used to determine the outer validation loss. This means that the outer validation loss indirectly depends on $x_p$. Also notice that $\theta^*$ has to be retrained for every candidate $x_p$. This results in two levels of optimization: an outer maximization (Equation \ref{eq:outer_optim}) and an inner minimization (Equation \ref{eq:inner_optim}). Hence, the problem of finding optimal poisoning points is a bilevel optimization problem. Due to the computational challenges in solving such a bilevel optimization problem \cite{biggio2012poisoning, cina2023wild,poisoning_attack_1,huang2020metapoison}, heuristic attack methods have emerged as appealing options \cite{cina2021hammer, kara2022beta}.

\subsection{Beta Poisoning Attack} \label{sec:BP}

Beta Poisoning, proposed by Cina et al.~\cite{cina2021hammer}, is a poisoning attack which aims to decrease the accuracy of ML models by injecting maliciously crafted poisoning samples into $\mathcal{D}_{tr}$. Instead of solving the aforementioned bilevel optimization problem, Beta Poisoning proposes a heuristic strategy. Its strategy is to poison the target distribution of $y_t$ with sample $x_p$ by maximizing the likelihood $P(x_p|y_t)$, making the training dataset linearly nonseparable. Formally, the optimization problem of Beta Poisoning can be stated as follows \cite{cina2021hammer}:
\begin{align}
\label{eq:likelihood}
&\argmax_{x_p} \qquad P(x_p|y_t) \\
&\;\text { s.t. } \qquad \mathbf{x}_{lb} \preceq x_{p} \preceq \mathbf{x}_{ub}
\end{align}
Note that this optimization is no longer bilevel, since there is no inner optimization. To estimate $P(x_p|y_t)$, Gaussian Kernel Density Estimator (KDE) is used, though alternative KDEs may also be considered.

Another feature of the Beta Poisoning attack is that it generates poisoning samples using linear combinations of other samples, called prototypes. Let $\mathcal{S} = \{ x_1, x_2, ..., x_k \}$ be the set of samples called the prototypes. For example, to construct $\mathcal{S}$, prototypes can be sampled randomly from $\mathcal{D}_{val}$. Given coefficients $\boldsymbol{\beta} \in \mathbb{R}^k$, poisoning sample $x_p$ is derived as:
\begin{equation}
\label{eq:adv_sample}
x_{p} = \psi(\boldsymbol{\beta}, \mathcal{S}) = \sum_{x_{i} \in \mathcal{S}} \boldsymbol{\beta}_{i} x_{i}
\end{equation}
Here, $\psi$ denotes the linear combination function. A key part of the Beta Poisoning attack is the values of the $\boldsymbol{\beta}$ coefficients. An optimization process is used to determine $\boldsymbol{\beta}$ coefficients. After $\boldsymbol{\beta}$ coefficients are determined, the final poisoned sample $x_p$ can be efficiently created via $\psi(\boldsymbol{\beta}, \mathcal{S})$. 

\begin{algorithm}[!t]
\hspace*{\algorithmicindent} \textbf{Input:} $\mathcal{D}_{val}$, $y_t$, $k$, $\mathbf{x}_{lb}$, $\mathbf{x}_{ub}$ \\
\hspace*{\algorithmicindent} \textbf{Output:} Poisoning sample $x_p$ 
\begin{algorithmic}[1]

\State $\mathcal{S}$ = \textsc{Sample\_Prototypes}($\mathcal{D}_{val}$, $y_t$, $k$)
\State $\boldsymbol{\beta}$ = \textsc{Initialize\_Beta}($k$)
\Repeat
    \State $x_p = \textsc{Clip}(\psi(\boldsymbol{\beta}, \mathcal{S}), \; \mathbf{x}_{lb}, \; \mathbf{x}_{ub})$
    \State $p = \textsc{Estimate} ~ P(x_p|y_t)$
    \State $\boldsymbol{\beta} = \boldsymbol{\beta} + \alpha \nabla_{\boldsymbol{\beta}} p$
\Until{stop condition is reached}

\State $x_p = \textsc{Clip}(\psi(\boldsymbol{\beta},\mathcal{S}), \; \mathbf{x}_{lb}, \; \mathbf{x}_{ub})$

\State \Return $x_p$
\end{algorithmic}
\caption{Pseudocode of Beta Poisoning}
\label{alg:beta_poisoning}
\end{algorithm}

The overall algorithm of the Beta Poisoning attack is given in Algorithm \ref{alg:beta_poisoning}. Its inputs are the validation set $\mathcal{D}_{val}$, the class targeted by the attack $y_t \in \mathcal{Y}$, the number of prototypes $k$, and the lower and upper bounds $\mathbf{x}_{lb}$, $\mathbf{x}_{ub}$. First, Algorithm \ref{alg:beta_poisoning} constructs the set of prototypes $\mathcal{S}$ by drawing $k$ random samples from $\mathcal{D}_{val}$, such that each sample belongs to class $y_t$ (line 1). Then, $\boldsymbol{\beta}$ coefficients are initialized by sampling from a uniform distribution between [0,1] (line 2). The main optimization of Beta Poisoning takes place between lines 3-7. On line 4, an initial $x_p$ is generated using the linear combination $\psi(\boldsymbol{\beta}, S)$ and clipped so that its feature values remain between $\mathbf{x}_{lb}$, $\mathbf{x}_{ub}$. Likelihood $P(x_p|y_t)$ is estimated using a Gaussian KDE on line 5. On line 6, based on the estimated likelihood $p$, the $\boldsymbol{\beta}$ coefficients are updated using gradient ascent. Here, $\alpha$ is the learning rate (by default, $\alpha = 0.01$). The optimization between lines 3-7 is executed repeatedly until the stop condition is met (line 7). Following \cite{cina2021hammer}, we use the stop condition that $P(x_p|y_t)$ should not change more than $1e-05$ in consecutive iterations. Finally, on lines 8-9, the poisoning sample $x_p$ is generated using the optimized $\boldsymbol{\beta}$ coefficients, clipped, and returned.

\section{Proposed Defenses} \label{sec:defenses}

In this section, we describe the defenses we propose for the Beta Poisoning attack. First, we provide our observations and defense insights, which form the starting points of our defenses. Then, we explain each defense one by one. 

\subsection{Observations and Insights} \label{sec:insights}

\begin{figure}[!t]
\centering
\begin{minipage}[b]{0.3\textwidth}
    \includegraphics[width=\textwidth]{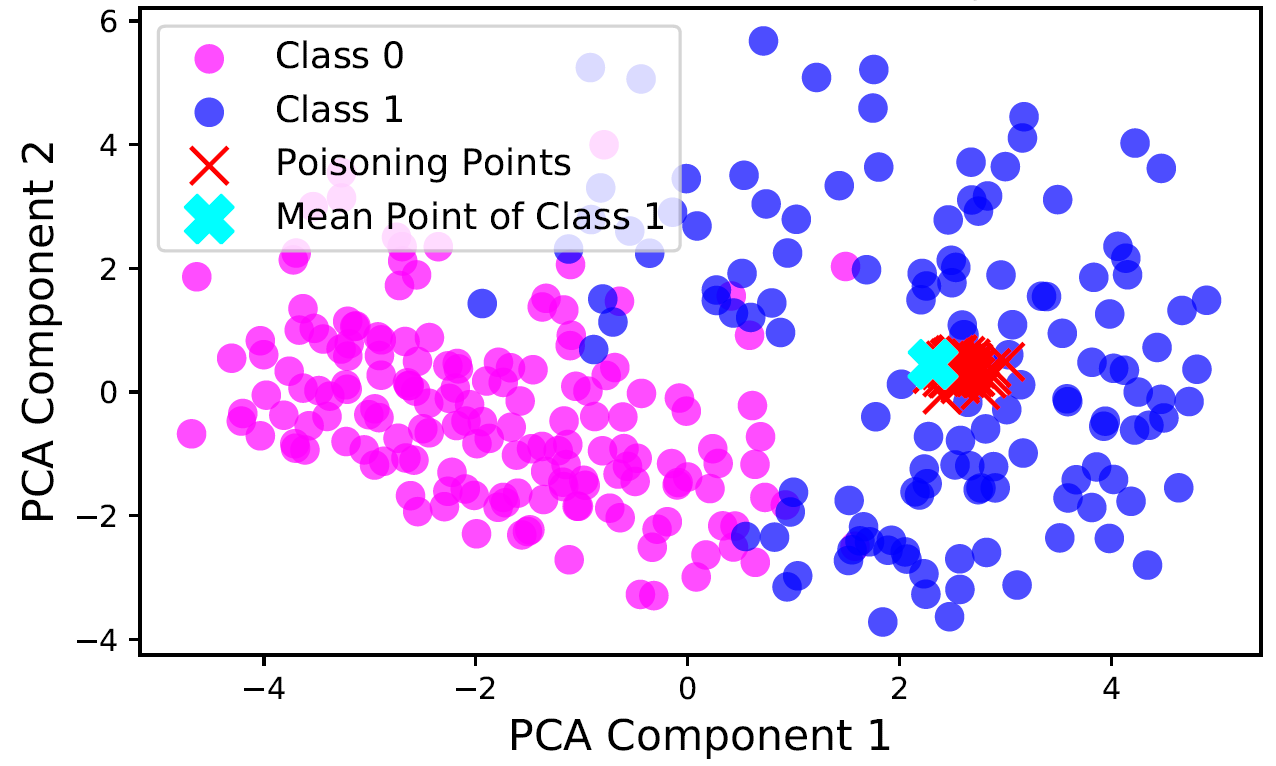}
\end{minipage}%
\caption{Visualization of actual and poisoning points from the MNIST dataset.}
\label{fig:MNIST_poisoned}
\end{figure}

To develop effective defenses for the Beta Poisoning attack, we first analyzed the distinguishing characteristics of the poisoning samples generated by the attack. We observed that the samples generated by the attack satisfy two properties:
\begin{itemize}
    \item Close proximity: Many of the poisoning samples have close proximity to one another. 
    \item Centered around the mean of $y_t$: The poisoning samples are near the mean of the target class $y_t$. 
\end{itemize}
We demonstrate these two properties in Figure \ref{fig:MNIST_poisoned}. The figure is drawn with the MNIST dataset (more information about the experiment setup and datasets are provided in Section \ref{sec:experiments}). We performed Principal Components Analysis (PCA) on the dataset and created a plot with two principal components. Each point represents a sample from the dataset. In addition to the actual samples from two classes (shown in pink and blue), the mean of the target class $y_t$ and the poisoning samples generated by the attack are also drawn. One can observe from the figure that the poisoning points are indeed located very close to one another. In addition, they are close to the mean of $y_t$, which is far from the mean of the non-target class. 

The two observations are intuitive. First, poisoning samples are generated via a linear combination $\psi(\boldsymbol{\beta}, \mathcal{S})$. Considering that the prototypes in $\mathcal{S}$ are representative of the actual data distribution, taking their linear combination results in samples that are close to the mean. Second, Algorithm \ref{alg:beta_poisoning} optimizes $\boldsymbol{\beta}$ by taking into account the estimation of $P(x_p|y_t)$. Samples which are located close to the mean would have higher $P(x_p|y_t)$ per Gaussian KDE (this would hold for other KDEs as well). Consequently, $\boldsymbol{\beta}$ is optimized in a way that favors the generation of samples close to the mean.

We leverage these observations when designing our defenses. Our defenses use the insights that the poisoning samples are tightly clustered with close proximity to one another, and this tight cluster occurs near the mean of the target class $y_t$, which is typically far from the mean of the non-target class. We aim to locate poisoning samples by finding such tightly distributed samples near the mean of the opposing class and far from the mean of the class that they claim to belong to.

\subsection{kNN Proximity-Based Defense (KPB)}

This defense utilizes the insight that poisoning samples tend to have significantly shorter distances to their neighbors compared to legitimate samples whose distances to their neighbors can be larger. To implement the defense, we use an approach based on k-nearest neighbors (kNN) \cite{cover1967nearest}. For each sample, we identify its nearest neighbors and compute that sample's average distance to its neighbors. If this distance is low, then the defense flags the sample as a likely poisoning sample. 

\begin{algorithm}[!t]
\hspace*{\algorithmicindent} \textbf{Input:} $\mathcal{D}_{sp}$, $\tau$, $\eta$ \\
\hspace*{\algorithmicindent} \textbf{Output:} $\mathcal{D}_{fl}$
\begin{algorithmic}[1]
\State $\mathcal{D}_{fl} \gets \emptyset$
\State $num \gets |\mathcal{D}_{sp}| \times \eta$ \;
\For{each sample $(x_i,y_i) \in \mathcal{D}_{sp}$}
\State $nbrs \gets \textsc{Identify\_Neighbors}(x_i, \mathcal{D}_{sp}, num)$ \;
\State $total\_dist \gets 0$
\For{each sample $(x_j,y_j) \in nbrs$}
\State $total\_dist \gets total\_dist + \textsc{dist}(x_i,x_j)$ \;
\EndFor
\State $avg\_dist \gets total\_dist / num$
\If{$avg\_dist < \tau$}
\State Insert $(x_i,y_i)$ into $\mathcal{D}_{fl}$
\EndIf
\EndFor
\State \Return $\mathcal{D}_{fl}$
\end{algorithmic}
\caption{Pseudocode of kNN Proximity-Based Defense}
\label{alg:knn_proximity}
\end{algorithm}

A more formal description of the defense is provided in Algorithm \ref{alg:knn_proximity}. The algorithm takes as input the suspicious dataset $\mathcal{D}_{sp}$, i.e., the dataset suspected of containing both legitimate training samples and poisoning samples. In addition, there are two more inputs: the $\tau$ parameter is used as the distance threshold, and the $\eta$ parameter is used to determine the number of neighbors. The output of Algorithm \ref{alg:knn_proximity} is $\mathcal{D}_{fl}$, i.e., the set of samples which are flagged by the algorithm as poisoning samples. At the beginning of the algorithm, $\mathcal{D}_{fl}$ is initialized as empty. The number of neighbors is determined by multiplying $\eta$ and the cardinality of $\mathcal{D}_{sp}$. Here, $\eta$ takes values between 0 and 1. By default, we use $\eta$ = 0.1. Determining the number of neighbors as a fraction of $\mathcal{D}_{sp}$'s cardinality is done to achieve better consistency across varying $\mathcal{D}_{sp}$ sizes. Then, for each sample $(x_i,y_i)$ in $\mathcal{D}_{sp}$, $x_i$'s $num$ number of nearest neighbors in $\mathcal{D}_{sp}$ are found by the \textsc{Identify\_Neighbors} function and stored in a list called $nbrs$. The average distance of $x_i$ to its $nbrs$ is computed between lines 5-9. If this average distance is lower than the $\tau$ threshold, then the current sample $(x_i,y_i)$ is flagged as a poisoning sample and inserted to $\mathcal{D}_{fl}$. 

\subsection{Neighborhood Class Comparison Defense (NCC)}

Recall from Figure \ref{fig:MNIST_poisoned} that poisoning samples are in close proximity to one another, near the mean of $y_t$. As such, for a poisoning sample, its close neighbors are likely to have the same class label $y$ as the poisoning sample itself. However, its relatively more distant neighbors are likely to have the opposite class label. For example, consider a poisoning sample denoted with red X symbol in Figure \ref{fig:MNIST_poisoned}. Its closest neighbors are also red X symbols, belonging to the same class as the poisoning sample. However, since the poisoning sample is near the mean of the blue class, it is surrounded by blue points, i.e., its relatively distant neighbors have the blue class label. This discrepancy of class labels between close neighbors versus relatively distant neighbors is the factor we use in our defense. 

\begin{algorithm}[!t]
\hspace*{\algorithmicindent} \textbf{Input:} $\mathcal{D}_{sp}$, $\eta$ \\
\hspace*{\algorithmicindent} \textbf{Output:} $\mathcal{D}_{fl}$
\begin{algorithmic}[1]
\State $\mathcal{D}_{fl} \gets \emptyset$
\State $num \gets |\mathcal{D}_{sp}| \times \eta$ \;
\For{each sample $(x_i,y_i) \in \mathcal{D}_{sp}$}
\State $nbrs_1 \gets \textsc{Identify\_Neighbors}(x_i, \mathcal{D}_{sp}, num)$ \;
\State $nbrs_2 \gets \textsc{Identify\_Neighbors}(x_i, \mathcal{D}_{sp}, 2 \times num)$ \;
\State $y_1 \gets \textsc{Majority\_Class}(nbrs_1)$ \;
\State $y_2 \gets \textsc{Majority\_Class}(nbrs_2)$ \;
\If{$y_1 \neq y_2$}
\State Insert $(x_i,y_i)$ into $\mathcal{D}_{fl}$ \;
\EndIf
\EndFor
\State \Return $\mathcal{D}_{fl}$
\end{algorithmic}
\caption{Pseudocode of NCC Defense}
\label{alg:ncc}
\end{algorithm}

The formal description of our NCC defense is provided in Algorithm \ref{alg:ncc}. The $\mathcal{D}_{sp}$ and $\eta$ inputs of the NCC defense are identical to our previous defense. For each sample $(x_i,y_i)$ in $\mathcal{D}_{sp}$, $x_i$'s $num$ number of nearest neighbors are found using the \textsc{Identify\_Neighbors} function. In addition, $x_i$'s $2 \times num$ number of nearest neighbors are also found by the \textsc{Identify\_Neighbors} function. They are stored in two sets called $nbrs_1$ and $nbrs_2$, respectively. Then, a majority vote is performed among the class labels of the samples in $nbrs_1$ and $nbrs_2$ separately, i.e., the most occurring class label in $nbrs_1$ is found and stored in $y_1$, and the most occurring class label in $nbrs_2$ is found and stored in $y_2$. If $y_1$ and $y_2$ are different, then the current sample $(x_i,y_i)$ is flagged as a poisoning sample.

\subsection{Clustering-Based Defense (CBD)}

This defense utilizes the insight that poisoning samples are near the mean of the target class $y_t$, which is typically far from the mean of the non-target class (denoted by $y_{nt}$). Thus, the distances between samples belonging to $y_{nt}$ are compared with the mean of $y_t$. Samples with small distances have higher chance of being poisoning samples. For example, we can observe from Figure \ref{fig:MNIST_poisoned} that the distances between poisoning samples and the mean of class 1 are indeed much smaller than the distances between pink samples and the mean of class 1. 

\begin{algorithm}[!t]
\hspace*{\algorithmicindent} \textbf{Input:} $\mathcal{D}_{sp}$ \\
\hspace*{\algorithmicindent} \textbf{Output:} $\mathcal{D}_{fl}$
\begin{algorithmic}[1]
\State $\mathcal{D}_{fl} \gets \emptyset$
\State $mean \gets \textsc{Compute\_Mean}(\mathcal{D}_{sp},y_{t})$
\State $distances \gets $ empty list
\For{$(x_i,y_i) \in \mathcal{D}_{sp}$ such that $y_i = y_{nt}$}
\State $distance \gets \textsc{dist}(x_i,mean)$
\State Insert tuple $((x_i,y_i),distance)$ to $distances$
\EndFor
\State Sort $distances$ according to $distance$ values 
\State $clusters \gets \textsc{Cluster}(distances)$
\State $min\_cluster \gets \textsc{Find\_Min\_Cluster}(clusters)$
\For{$(x_i,y_i) \in min\_cluster$}
\State Insert $(x_i,y_i)$ into $\mathcal{D}_{fl}$
\EndFor
\State \Return $\mathcal{D}_{fl}$
\end{algorithmic}
\caption{Pseudocode of Clustering-Based Defense}
\label{alg:clustering}
\end{algorithm}

The formal description of our clustering-based defense is provided in Algorithm \ref{alg:clustering}. The algorithm starts by computing the mean of $y_t$ using $\mathcal{D}_{sp}$. Here, we highlight an important design decision regarding why we choose to perform comparisons with the mean of $y_t$ but not $y_{nt}$, since at first sight, it can be thought that poisoning samples are distant from the mean of $y_{nt}$. The reason why we make this choice is because the mean of $y_{nt}$ is actually affected by the poisoning samples that are added to $\mathcal{D}_{sp}$, which have $y_{nt}$ labels. On the other hand, the mean of $y_t$ is unaffected by the Beta Poisoning attack. Thus, it is more reliable to use the mean of $y_t$. 

After the mean of $y_t$ is found, the list of tuples named $distances$ is initialized as empty. For each sample in $\mathcal{D}_{sp}$ which belongs to the target class $y_{nt}$, we compute the distance between that sample and the mean of $y_t$. This distance is inserted into $distances$. After the insertions are complete, $distances$ are sorted in ascending order and clustered. For clustering, we use the well-known k-means clustering algorithm \cite{kanungo2002efficient}. The Elbow method is used to determine the optimal number of clusters in k-means \cite{hastie2009elements}, which eliminates the need for an additional parameter for the number of clusters. After clustering (line 9), the cluster with the smallest distances is found (line 10). Samples in this cluster are flagged as poisoning samples and inserted into $\mathcal{D}_{fl}$.

\subsection{Mean Distance Threshold Defense (MDT)} 

This defense shares a similar intuition to CBD, i.e., samples belonging to class $y_{nt}$ which have small distances to the mean of $y_t$ are likely to be poisoning samples. Instead of the clustering-based approach in CBD, this defense uses a threshold-based approach. Those samples belonging to $y_{nt}$ with distances smaller than a threshold to the mean of $y_t$ are flagged as poisoning samples. Although this approach is simpler, it is empirically effective.

\begin{algorithm}[!t]
\hspace*{\algorithmicindent} \textbf{Input:} $\mathcal{D}_{sp}$, $\tau$ \\
\hspace*{\algorithmicindent} \textbf{Output:} $\mathcal{D}_{fl}$
\begin{algorithmic}[1]
\State $\mathcal{D}_{fl} \gets \emptyset$
\State $mean \gets \textsc{Compute\_Mean}(\mathcal{D}_{sp},y_t)$
\For{$(x_i,y_i) \in \mathcal{D}_{sp}$ such that $y_i = y_{nt}$}
\State $distance \gets \textsc{dist}(x_i,mean)$
\If{$distance < \tau$}
\State Insert $(x_i,y_i)$ into $\mathcal{D}_{fl}$
\EndIf
\EndFor
\State \Return $\mathcal{D}_{fl}$
\end{algorithmic}
\caption{Pseudocode of MDT Defense}
\label{alg:mdt}
\end{algorithm}

The formal description of the defense is provided in Algorithm \ref{alg:mdt}. Similar to Algorithm \ref{alg:clustering}, the defense starts by computing the mean of samples in $y_t$. Then, for each sample in $\mathcal{D}_{sp}$ which belongs to the target class $y_{nt}$, the distance between that sample and the mean of $y_t$ is computed. If this distance is smaller than the threshold $\tau$, the current sample is flagged as a poisoning sample and inserted into $\mathcal{D}_{fl}$. 

\section{Experimental Evaluation} \label{sec:experiments} 

\subsection{Experiment Setup} 

\textbf{Datasets.} We evaluated our defenses using two datasets: MNIST and CIFAR-10. The MNIST dataset \cite{lecun1998gradient} contains a collection of grayscale images of handwritten digits, with each image possessing a resolution of 28x28 pixels. Each pixel has an intensity value ranging from 0 to 255. The dataset consists of 10 classes, each representing one digit, ranging from 0 to 9. The CIFAR-10 dataset \cite{krizhevsky2009learning} contains RGB images representing a variety of objects, such as airplanes, cars, birds, cats, and dogs. Images in this dataset have dimensions of 32x32 pixels. Similar to MNIST, CIFAR-10 contains 10 classes.

\textbf{Attack implementation and parameters.} We used the original implementation of the Beta Poisoning attack provided by the authors \cite{cina2021hammer} with default parameters. We used a poison rate of 20\% in order to have a sufficiently large number of poisoning samples. Following the experimental setup of \cite{cina2021hammer,kara2022beta}, we utilized a binary classification problem and selected the same classes for poisoning. Classes 4 and 6 were selected for MNIST, while classes 0 (airplane) and 8 (ship) were selected for CIFAR-10. For dimensionality reduction and visualization, we utilized Principal Components Analysis (PCA) \cite{gewers2021principal}. To improve statistical significance and reliability of our results, we repeated each experiment 5 times and averaged the results.

\textbf{Metrics.} We use well-known metrics such as precision, accuracy, recall, and F1-score to measure the effectiveness of our defenses. Before we formalize these metrics, we define True Positives (TP), False Positives (FP), True Negatives (TN), and False Negatives (FN) in our defenses' context as follows. Let $(x_i,y_i) \in \mathcal{D}_{sp}$ denote a sample. Then: 
\begin{itemize}
    \item True Positive (TP): $(x_i,y_i)$ was generated by the Beta Poisoning attack, and the defense correctly flagged it as a poisoning sample, i.e., $(x_i,y_i)$ was included in $\mathcal{D}_{fl}$.
    \item False Positive (FP): $(x_i,y_i)$ was a legitimate sample not generated by the Beta Poisoning attack, but the defense incorrectly flagged it as a poisoning sample, i.e., $(x_i,y_i)$ was included in $\mathcal{D}_{fl}$.
    \item True Negative (TN): $(x_i,y_i)$ was a legitimate sample not generated by the Beta Poisoning attack, and the defense did not include it in $\mathcal{D}_{fl}$.
    \item False Negative (FN): $(x_i,y_i)$ was generated by the Beta Poisoning attack, but the defense failed to include it in $\mathcal{D}_{fl}$.
\end{itemize}
Following these definitions, accuracy, precision, recall, and F1-score are defined as:
\begin{equation}
    \text{Accuracy} = \frac{\text{TP + TN}}{\text{TP + TN + FP + FN}}
\end{equation}
\begin{equation}
    \text{Precision} = \frac{\text{TP}}{\text{TP +  FP}} \qquad \text{Recall} = \frac{\text{TP}}{\text{TP + FN}}
\end{equation}
\begin{equation}
    \text{F1-Score} = \frac{2 \times \text{Precision} \times \text{Recall}}{\text{Precision + Recall}} 
\end{equation}

\subsection{Comparison of Defenses}

We start by comparing our four defenses side by side. In this comparison, we compare the best-performing versions of the defenses, i.e., the parameters in each defense are optimized individually before the comparison, and the parameters which maximize defense accuracy are selected. 

\begin{table*}[t]
\renewcommand{\arraystretch}{1.15}
\centering
\caption{\small Comparison of Defenses on MNIST and CIFAR-10 Datasets}
\resizebox{\textwidth}{!}{
\begin{tabular}{|c|c|c|c|c|c|c|c|c|c|c|} 
\cline{2-9}
\multicolumn{1}{c|}{} & \multicolumn{2}{c|}{kNN Proximity-Based} & \multicolumn{2}{c|}{Neighborhood Comparison} & \multicolumn{2}{c|}{Clustering-Based} & \multicolumn{2}{c|}{Mean Distance Threshold}\\ 
\hline
& MNIST & CIFAR-10 & MNIST & CIFAR-10 & MNIST & CIFAR-10 & MNIST & CIFAR-10           \\ 
\hline
Accuracy  & 1.0 & 1.0 & 0.830 & 0.897 & 1.0 & 0.992 & 1.0 & 1.0   \\ 
\hline
F1-Score  & 1.0 & 1.0 & 0.682 & 0.764 & 1.0 & 0.976 & 1.0 & 1.0  \\
\hline
Precision  & 1.0 & 1.0 & 0.545 & 0.619 & 1.0 & 0.952 & 1.0 & 1.0   \\
\hline
Recall   & 1.0 & 1.0 & 1.0 & 0.980 & 1.0 & 1.0 & 1.0 & 1.0   \\
\hline
\end{tabular}
}
\label{tab:comparison_defenses}
\end{table*}

Table \ref{tab:comparison_defenses} shows the results of applying our defenses on MNIST and CIFAR-10 datasets. The results indicate that both the kNN Proximity-Based Defense (KPB) and the Mean Distance Threshold Defense (MDT) are able to reach accuracy, F1-score, precision, and recall values equal to 1.0. In contrast, although the two remaining defenses can reach high recall values (1.0 or 0.98), their precisions may suffer. The fact that they have high recall but relatively lower precision, which also fuels decreased accuracy and F1-scores, means that these defenses yield false positives. The number of false positives seems to be especially high in the Neighborhood Class Comparison Defense (NCC). We believe that this is because there can also be many legitimate samples for which the $num$ nearest neighbors have a different majority class compared to the $2 \times num$ nearest neighbors. This holds especially true for legitimate samples that are close to the decision boundary. Fine-tuning the defense to address such samples can be a good direction for future work.

In general, two of our defenses (KPB and MDT) reach perfect 1.0 scores in all metrics and on both datasets. CBD also reaches perfect 1.0 scores on the MNIST dataset and close to 1.0 scores ($\geq$ 0.95) on the CIFAR-10 dataset. Only the NCC defense lags behind, mostly due to its precision. Overall, we can conclude that our defenses are quite effective in defending against Beta Poisoning. 

\subsection{Experiments with Individual Defenses}

Next, to gain deeper insights into the four defenses, we conducted experiments with each defense individually.

\begin{figure}[!t]
\centering
\begin{minipage}[b]{0.47\textwidth}
    \includegraphics[width=\textwidth]{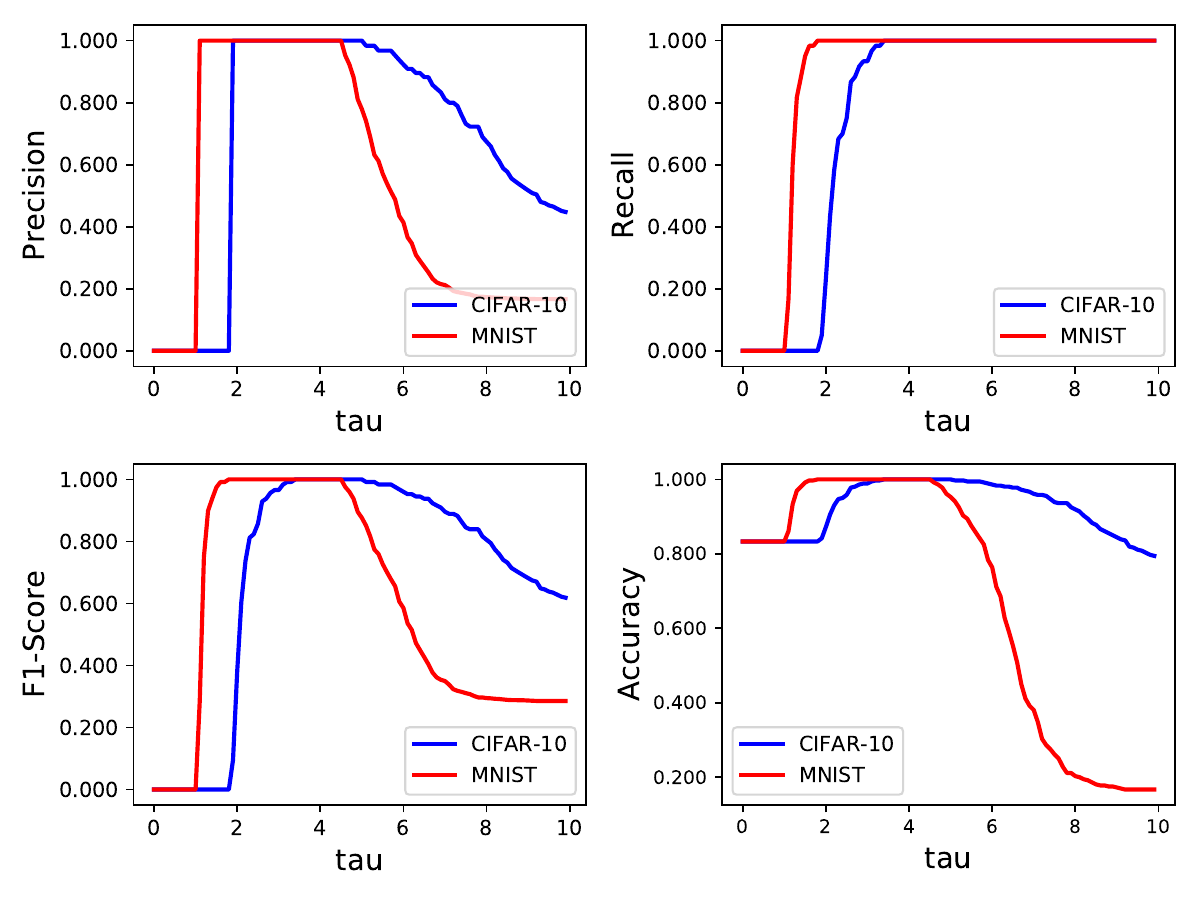}
\end{minipage}%
\vspace{-6pt}
\caption{Impact of $\tau$ on the results of the kNN Proximity-Based (KPB) defense.}
\label{fig:KPB_results}
\vspace{-4pt}
\end{figure}

\textbf{Experiments with KPB.} We start with the KPB defense. In KPB, choosing the right threshold $\tau$ is important. Therefore, we perform experiments with varying $\tau$ and measure the changes in performance metrics for the MNIST and CIFAR-10 datasets. Results are shown in Figure \ref{fig:KPB_results}. 

Based on the results, the best $\tau$ values for both datasets seem to be between 3 and 5. For these $\tau$ values, both the accuracy and F1-scores of the defense reach 1.0, which means that all samples are correctly predicted. However, when $\tau$ is smaller than 3, the precision and recall of the defense drop significantly. This shows that low values of $\tau$ are unable to identify poisoning samples effectively. On the other hand, when $\tau$ is large, e.g., larger than 5, although the recall remains high, precision starts decreasing. This shows that large $\tau$ yields many false positives. The increased number of false positives also causes accuracy and F1-scores to decrease. Overall, these results show that $\tau$ should be neither too large nor too small.

Another interesting aspect is to study the impacts of the different datasets. The two datasets seem to show similar trends; however, due to the nature of their content, the actual $\tau$ value which provides best results may differ from dataset to dataset. For the MNIST and CIFAR-10 datasets, we observe that both datasets favor $\tau$ between 3 and 5. However, more generally, we expect that identifying poisoning points with KPB becomes challenging in datasets where samples are more densely distributed. Consequently, the effectiveness of KPB may fluctuate depending on the distribution of the dataset. Thus, slightly different yet consistent behavior on CIFAR-10 and MNIST is intuitive.

\begin{figure}[!t]
\centering
\begin{minipage}[b]{0.47\textwidth}
    \includegraphics[width=\textwidth]{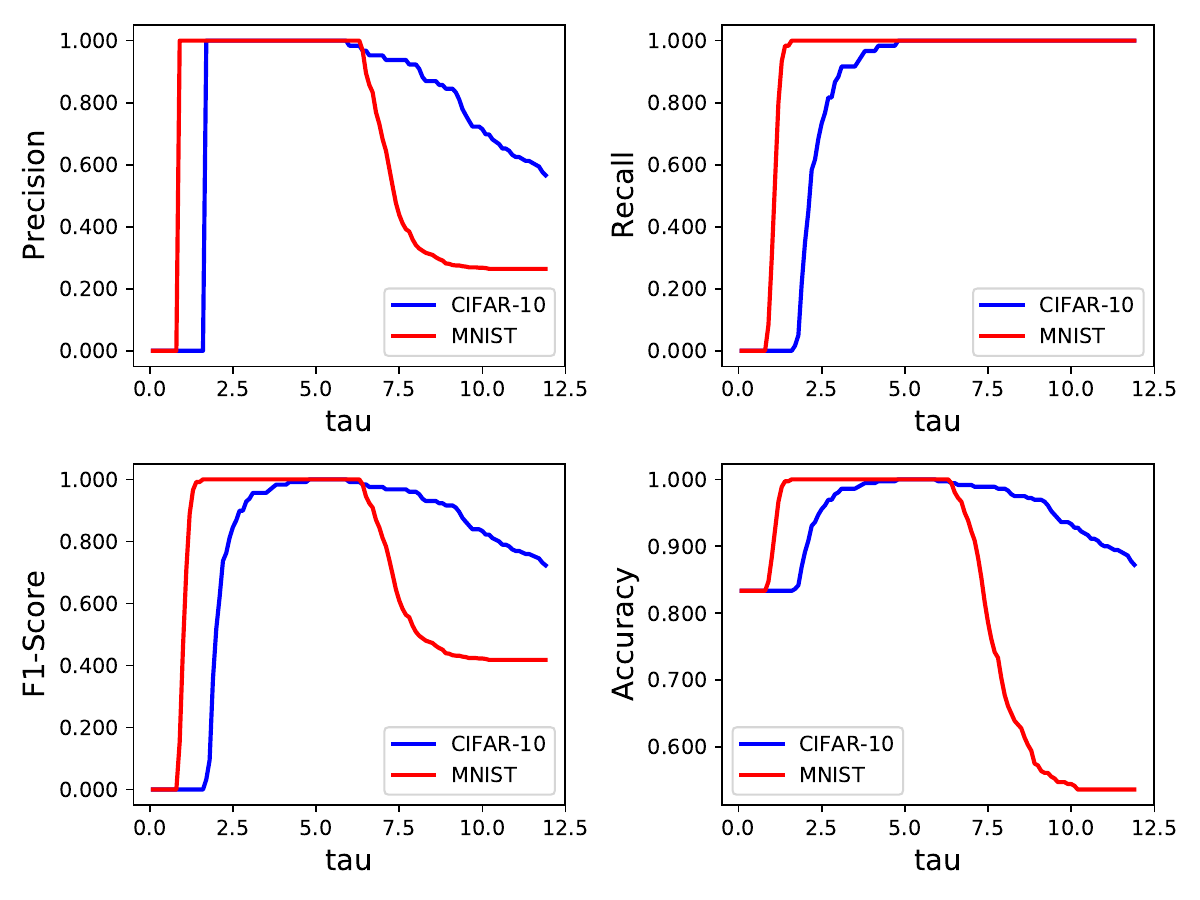}
\end{minipage}%
\vspace{-6pt}
\caption{Impact of $\tau$ on the results of the Mean Distance Threshold (MDT) defense.}
\label{fig:MDT_results}
\vspace{-4pt}
\end{figure}

\textbf{Experiments with MDT.} MDT also uses a distance threshold parameter $\tau$. Thus, similar to KPB, we perform experiments with varying $\tau$ in MDT and measure the changes in performance metrics for the MNIST and CIFAR-10 datasets. Results are shown in Figure \ref{fig:MDT_results}.

According to the results, the best $\tau$ values for both datasets are between 5 and 7. The accuracy and F1-scores of the defense reach 1.0 when $\tau$ is selected this way. Similar to KPB, when $\tau$ is smaller than this ideal range, the precision and recall of the defense drop. Also, when $\tau$ is larger than this range, recall remains high, but precision starts decreasing, yielding decreased F1-score and accuracy as well. Overall, we arrive at a conclusion which is similar to KPB: value of the $\tau$ parameter should be neither too large nor too small. 

We also observe from Figure \ref{fig:MDT_results} that large $\tau$ causes stark decrease in performance on the MNIST dataset, whereas the performance decrease is more gradual and mild on the CIFAR-10 dataset. This behavior is consistent with the behavior in Figure \ref{fig:KPB_results}. The distribution of samples in CIFAR-10 is more mixed (i.e., samples from opposing classes have closer distances to one another), whereas the classes in MNIST are distinct, i.e., samples from opposing classes are far away. As such, increased $\tau$ has a limited impact on MNIST for a while, but after a certain point, it has a stark impact. In contrast, the impact is more gradual on CIFAR-10. 

\begin{figure}[t]
\centering
\begin{minipage}[b]{0.237\textwidth}
    \includegraphics[width=\textwidth]{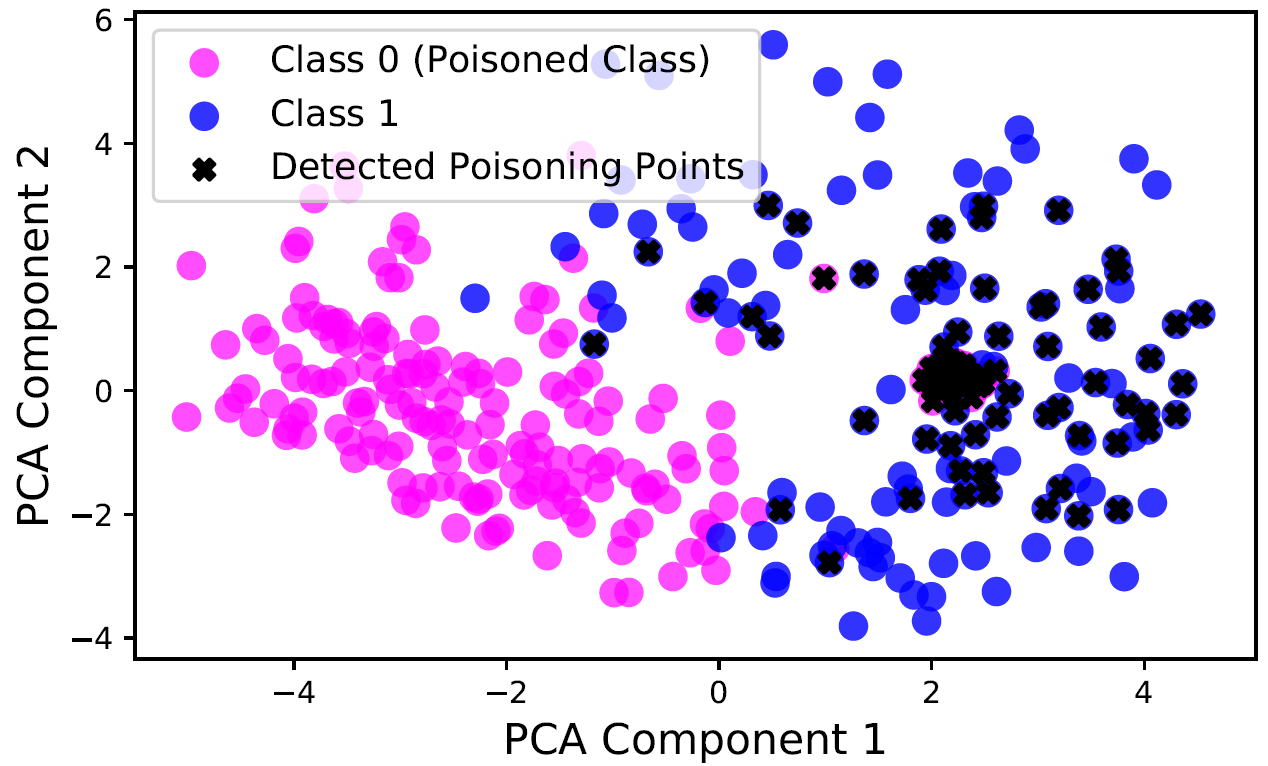}
\end{minipage}%
~~~
\begin{minipage}[b]{0.237\textwidth}
    \includegraphics[width=\textwidth]{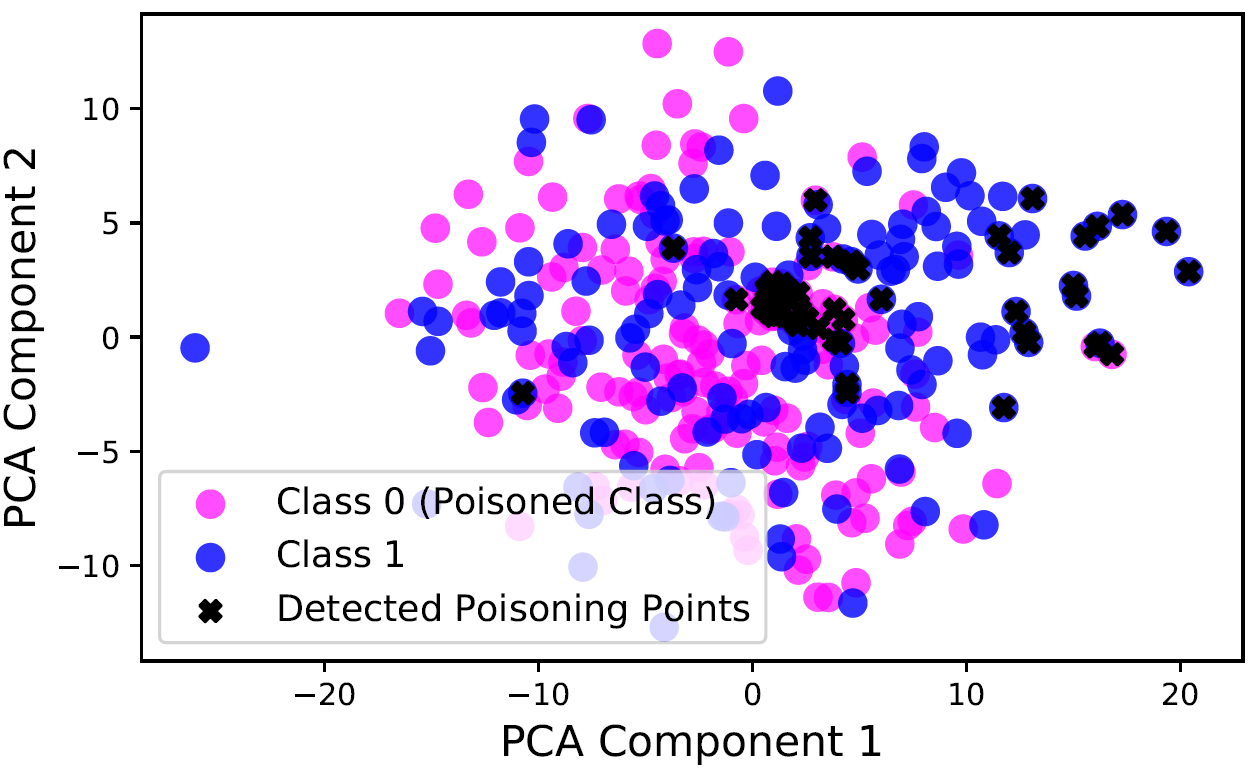}
\end{minipage}%
\vspace{-4pt}
\caption{Legitimate versus poisoning points detected by the NCC defense (MNIST dataset on the left, CIFAR-10 dataset on the right).}
\label{fig:NCC_results}
\vspace{-2pt}
\end{figure}

\begin{figure}[t]
\centering
\begin{minipage}[b]{0.237\textwidth}
    \includegraphics[width=\textwidth]{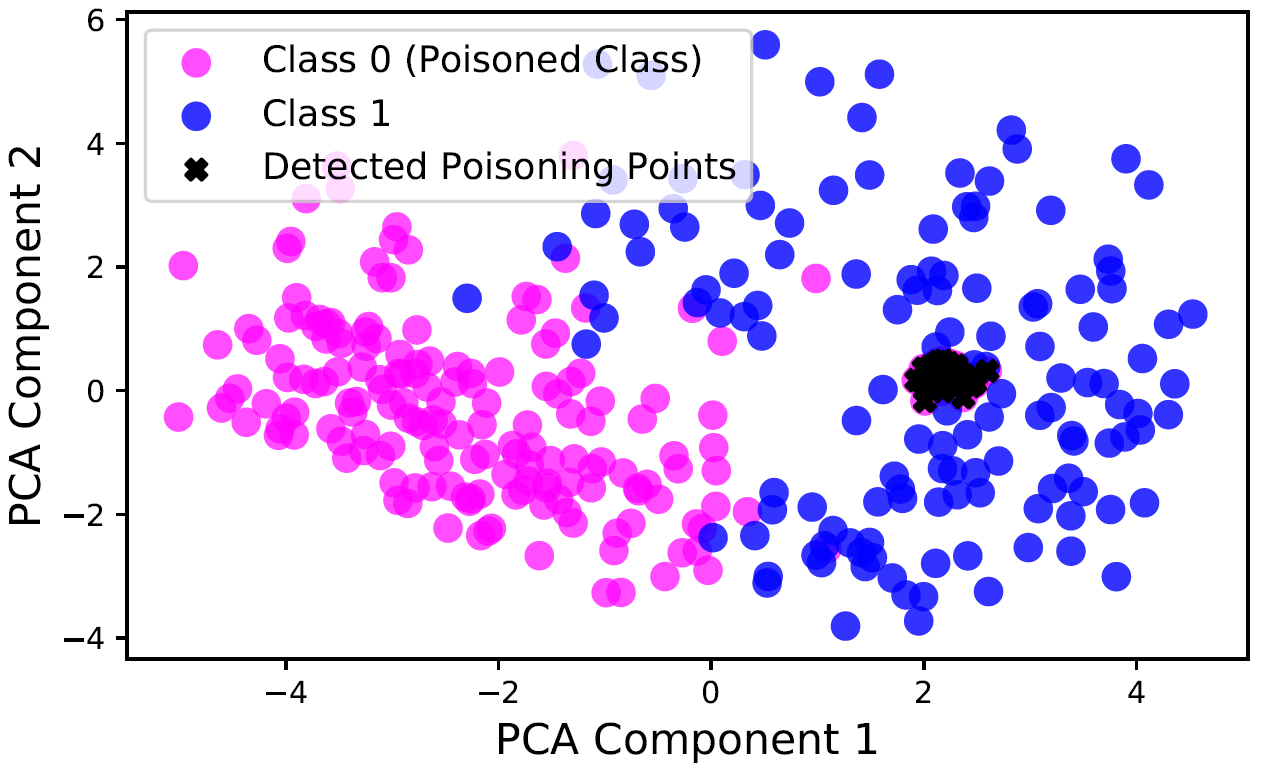}
\end{minipage}%
~~~
\begin{minipage}[b]{0.237\textwidth}
    \includegraphics[width=\textwidth]{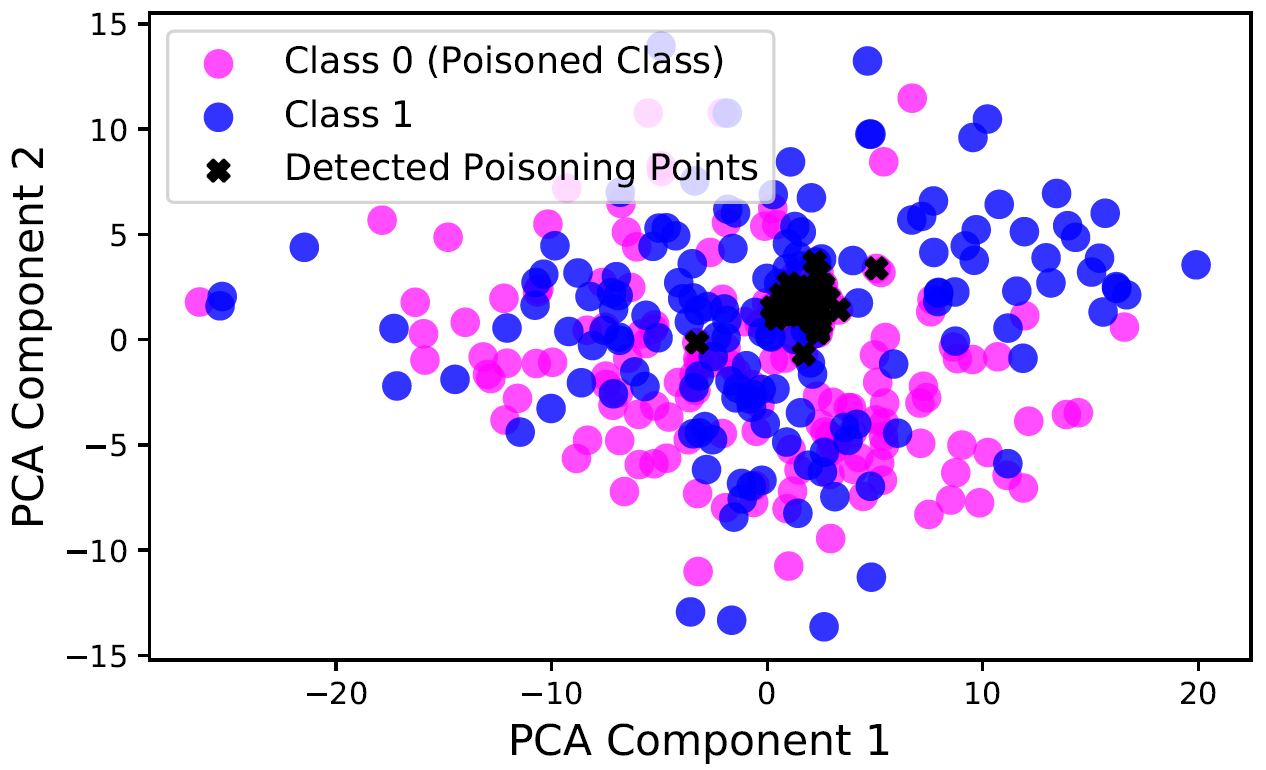}
\end{minipage}%
\vspace{-4pt}
\caption{Legitimate versus poisoning points detected by the CBD defense (MNIST dataset on the left, CIFAR-10 dataset on the right).}
\label{fig:CBD_results}
\vspace{-4pt}
\end{figure}

\textbf{Experiments with NCC and CBD.} For NCC and CBD, we visualize legitimate samples and detected poisoning samples using PCA. The results are shown in Figure \ref{fig:NCC_results} for NCC and in Figure \ref{fig:CBD_results} for CBD.  

NCC was developed based on the intuition that poisoning samples are closely located (clustered) together. According to Table \ref{tab:comparison_defenses}, NCC effectively identifies poisoning samples in MNIST and CIFAR-10, as implied by the high recall values. Nevertheless, the relatively lower precision values of the defense highlight the occurrence of false positives. We indeed observe this behavior in Figure \ref{fig:NCC_results}. There are several points colored in black which are not poisoning points; furthermore, they are far from the mean of the blue class. They typically occur in the relatively sparse regions of the space or near the samples of the pink class.

In CBD, samples' distances to the mean of $y_t$ are important. It can be observed from Figure \ref{fig:CBD_results} that, indeed, the samples which are predicted as poisoning points are very near the mean of the blue class. The visualizations in Figure \ref{fig:CBD_results} also show correlations with the results in Table \ref{tab:comparison_defenses}. CBD achieves perfect scores on the MNIST dataset according to Table \ref{tab:comparison_defenses} and the MNIST visualization in Figure \ref{fig:CBD_results} supports that because of the clear separability between pink and blue classes. Since the two classes are separable, any pink point near the mean of the blue class is easily caught as a poisoning sample by the defense. In contrast, Table \ref{tab:comparison_defenses} shows that the defense performs worse on the CIFAR-10 dataset. This is a reasonable outcome according to Figure \ref{fig:CBD_results} since the two classes are not easily separable. Thus, a point which is near the mean of the blue class may also legitimately be a member of the pink class but not a poisoning point. CBD is likely to yield a false positive result for such a point.

\subsection{Preliminary Experiments with CIFAR-100}

In order to explore how our defenses perform in more challenging scenarios (e.g., datasets with higher complexity), we also performed preliminary experiments with the CIFAR-100 dataset. The results are provided in Table \ref{tab:cifar100}. It can be observed that the results in Table \ref{tab:cifar100} are parallel to the results in Table \ref{tab:comparison_defenses}, but the defenses perform slightly worse on CIFAR-100 compared to the earlier datasets, due to the increased complexity of CIFAR-100. Similar to previous results, KPB, CBD, and MDT defenses provide good results (e.g., high accuracy and F1-scores) whereas NCC lags behind. It can again be observed that this is primarily caused by NCC's low precision. Overall, we can conclude that our defenses usually achieve strong accuracy and F1 scores (greater than 0.99 and 0.98, respectively) on CIFAR-100 as well.

\begin{table}[t]
\renewcommand{\arraystretch}{1.15}
\centering
\caption{Comparison of Defenses on CIFAR-100}
\begin{tabular}{|c|c|c|c|c|c|c|} 
\cline{2-5}
\multicolumn{1}{c|}{} & \multicolumn{1}{c|}{KPB} & \multicolumn{1}{c|}{NCC} & \multicolumn{1}{c|}{CBD} & \multicolumn{1}{c|}{MDT}\\ 
\hline
Accuracy  & 0.994 & 0.747 & 0.997 & 0.994 \\ 
\hline
F1-Score  & 0.983 & 0.556 & 0.992 & 0.983 \\
\hline
Precision  & 1.0 & 0.393 & 1.0 & 0.983 \\
\hline
Recall   & 0.966 & 0.950 & 0.983 & 0.983  \\
\hline
\end{tabular}
\label{tab:cifar100}
\end{table}

\section{Conclusion} \label{sec:conclusion}

Poisoning attacks pose a serious threat to ML security, with Beta Poisoning being one recent attack that disrupts model accuracy by making the training dataset linearly nonseparable. In this paper, we proposed four specialized defense strategies targeting Beta Poisoning attacks: KPB, NCC, CBD, and MDT, leveraging key observations regarding poisoning samples, such as their clustering tendencies and proximity to the mean of target and non-target classes. Evaluations on MNIST and CIFAR-10 datasets showed that all defenses achieve strong accuracy and F1-scores. Notably, KPB and MDT are shown to achieve perfect scores, highlighting their effectiveness.

There are several avenues for future work. First, exploring the effectiveness of our defenses on a wider range of datasets and models would help assess their generalizability. Second, combining our defenses with other defenses from the literature (e.g., defenses specialized in defending against other poisoning attacks) can be considered. Third, although our defenses target Beta Poisoning attacks, it would be interesting to study whether our defenses are effective against other types of poisoning attacks as well. Finally, exploring attack strategies specifically optimized to evade the proposed defenses would be an interesting aspect.

\bibliographystyle{IEEEtran}
\bibliography{references.bib}

\end{document}